\documentclass[nopreprintline,12pt,authoryear]{elsarticle}

\usepackage{amssymb}

\usepackage{multirow}
\journal{RBE Journal}

\begin{document}

\begin{frontmatter}

\title{A Deep Convolutional Neural Network for COVID-19 Detection Using Chest X-rays}

\author{Pedro R. A. S. Bassi, Romis Attux}

\address{Department of Computer Engineering and Industrial Automation, School of Electrical and Computer Engineering, University of Campinas - Campinas, Brazil p157007@dac.unicamp.br, ORCID: 0000-0002-8995-9423, attux@dca.fee.unicamp.br }

\begin{abstract}

Purpose:
We present image classifiers based on Dense Convolutional Networks and transfer learning to classify chest X-ray images according to three labels: COVID-19, pneumonia and normal.

Methods:
We fine-tuned neural networks pretrained on ImageNet and applied a twice transfer learning approach, using NIH ChestX-ray14 dataset as an intermediate step. We also suggested a novelty called output neuron keeping, which changes the twice transfer learning technique. In order to clarify the modus operandi of the models, we used Layer-wise Relevance Propagation (LRP) to generate heatmaps.

Results:
We were able to reach test accuracy of 100\% on our test dataset. Twice transfer learning and output neuron keeping showed promising results improving performances, mainly in the beginning of the training process. Although LRP revealed that words on the X-rays can influence the networks predictions, we discovered this had only a very small effect on accuracy.

Conclusion:
Although clinical studies and larger datasets are still needed to further ensure good generalization, the state-of-the-art performances we achieved show that, with the help of artificial intelligence, chest X-rays can become a cheap and accurate auxiliary method for COVID-19 diagnosis. Heatmaps generated by LRP improve the interpretability of the deep neural networks and indicate an analytical path for future research on diagnosis.  Twice transfer learning with output neuron keeping improved performances.

\end{abstract}





\end{frontmatter}


\section{Introduction}

\label{}

In 2020, COVID-19 became pandemic, affecting both developed and developing countries around the world. By October 2020, the virus had already infected more than 40,000,000 people and caused more than one million deaths (\cite{numbersCorona}).

The most commonly used method for COVID-19 diagnosis is reverse transcriptase-polymerase chain reaction (RT-PCR) (\cite{test}). It has a high specificity, but is also expensive, slow and currently at a high demand. Chest X-rays are commonly available and are faster and cheaper, but signals associated with the presence of COVID-19 in the lungs can be hard to detect.

Researchers have already suggested the use of deep neural networks (DNNs) to help in the detection of the disease on Chest X-ray images (\cite{CovidNet}, \cite{review}). In \cite{CovidNet}, the authors achieved good results, with 92.6\% test accuracy, 96.4\% recall and 87\% precision on the COVID-19 images. 

Deep neural networks (DNNs) have been successful at identifying pneumonia from X-rays, performing better than radiologists (\cite{chexnet}). In this work, we used Dense Convolutional Networks or DenseNets (\cite{DenseNet}). The first network is CheXNet (\cite{chexnet}), a 121 layers dense network (or DenseNet121) that had already been pretrained on ImageNet (\cite{imagenet}) and on NIH ChestX-ray14 dataset (\cite{NIHSet}), a database with over 100,000 frontal X-ray images, which contain 14 different diseases and also healthy individuals. We applied transfer learning to teach the neural network to differentiate between normal lungs, COVID-19 and pneumonia. Our COVID-19 dataset was assembled merging COVID-19, pneumonia and normal chest X-ray open datasets. Other DNN is a 201 layers DenseNet that had been pretrained on ImageNet (\cite{imagenet}) and we also fine tuned it in our COVID-19 database. 

CheXNet is a neural network that had been trained twice (on ImageNet and ChestX-ray14) and we trained it again, making our process a twice transfer learning or transfer learning in three steps. Inspired by this, we decided to explore this technique. We downloaded a 201-layer DenseNet, already pretrained on ImageNet, trained it on NIH ChestX-ray14 dataset (\cite{NIHSet}) and then on our smaller dataset containing the COVID-19 class. 

In this paper, we also suggest an original modification to twice transfer learning, which we called ``output neuron keeping". As the NIH ChestX-ray14 database already had the healthy and pneumonia classes, we suggest keeping the DNN output neurons for these classes in the last step of twice transfer learning (training on the COVID-19 dataset). Our hypothesis is that this will enable us to keep more of the information learned in the second dataset (ChestX-ray14) throughout training in the COVID-19 database, and in the final network. We tested this approach with another DenseNet201 and a CheXNet (maintaining only the neuron that classified pneumonia in this last case).

After training the DNNs, we applied Layer-wise Relevance Propagation (LRP) (\cite{LRP}), generating heatmaps of the X-rays, along with the probabilities of COVID-19, pneumonia and healthy lungs. These heatmaps show us the regions of the image that mostly influenced the network classification, and also regions that were more representative of other classes. Therefore, LRP can allow us to understand what the DNN finds relevant in an input image. But it can also be useful for a radiologist in identifying the effects of COVID-19 in the X-ray. An application of LRP in the context of neuroimaging can be seen at \cite{fmri-lrp}.

\section{Databases}

\subsection{NIH ChestX-ray14}
ChestX-ray14 is one of the largest chest X-ray datasets, with 112,120 images from 30,805 patients. The images were obtained by the US National Institutes of Health. It has 14 different diseases and also images with no findings. Some X-rays may show multiple conditions, making the classification of this dataset a multi-label classification problem. The database images originally had associated radiological reports, which were analyzed by the dataset authors, using natural language processing, to create automated labels for the X-rays. These labels, which will be used in this work, have an estimated accuracy higher than 90\% (\cite{NIHSet}). The dataset is unbalanced. State-of-the-art pneumonia-detecting DNNs were trained in this database: as an example, we can cite CheXNet (\cite{chexnet}), a DenseNet with 121 layers. 

\subsection{COVID-19 Database}

This database was assembled by the research group, joining images from 3 chest X-ray datasets, the first containing COVID-19 images, the second containing pneumonia images, and the third with healthy lungs images. 

We obtained 439 COVID-19 frontal X-rays from the dataset ``Covid-19 image data collection (\cite{GitCovidSet})'', downloaded on October 2020. To select these images we started with all COVID-19 frontal chest X-rays from \cite{GitCovidSet} and excluded images with arrows. This dataset was the largest COVID-19 X-ray collection that we could find. Is is also one of the best documented datasets, many images contained information about the patient age, gender, disease severity and image source. Their images were collected from public sources or indirectly from hospitals and physicians (\cite{GitCovidSet}).

The pneumonia images were extracted from the  CheXpert database. It is a collection of 224316 chest X-rays, from 65240 patients, classified according as 13 lung diseases or healthy (\cite{irvin2019chexpert}). We obtained 1255 images classified as pneumonia and  whose patients were older then 18 years old. Because the reported youngest COVID-19 patient in our dataset was 20 years old, we decided that keeping children in the other classes could be a source a bias. Like in NIH ChestX-ray14, all but 8 pneumonia images in this dataset were automatically labeled by the database authors, using natural language processing to  analyze radiology reports. These labels have an estimated accuracy that is higher then 90\%. 8 pneumonia images were manually labeled by three board-certified radiologists, these images were on the original CheXpert test dataset. The images were obtained in the Stanford University Hospital. 

The healthy images were obtained from the Montgomery and Shenzen databases (\cite{MontShen}). They are a collection of tuberculosis and healthy frontal chest X-rays, labeled by radiologists. The images were obtained from the Department of Health and Human Services, Montgomery County, Maryland, USA and Shenzhen No. 3 People’s Hospital in China (\cite{MontShen}). To select our images we separated all normal X-rays and removed the ones whose patients were underage. Therefore, we ended up with 370 normal images.

The labels that we used in our COVID-19 database had been created by the authors of the above mentioned datasets and were provided along with the X-rays. Therefore, the ``healthy'' and ``COVID-19'' labels were created by the datasets' authors) in accordance with radiological reports that accompanied the images, while most of the pneumonia labels were produced by using natural language processing to analyze radiological reports. The exceptions are 8 pneumonia labels that were manually created by three radiologists (we included these images in our test dataset, which will be explained in section ``Data processing and Augmentation'').

\subsection{Additional details of the assembled COVID-19 database}

In this section we will describe some general information about the patients in our COVID-19 database.

There were 370 healthy patients (the database from \cite{MontShen} does not contain multiple images from the same patient, according to the authors). The patients are 61.9\% male and their mean age is 36.1 years, with a standard deviation of 12.3 years.

Our pneumonia images were created with 1047 different patients. They are 58.2\% male, with a mean age of 61.8 years and an age standard deviation of 19.3 years.

Our COVID-19 database contained 268 patients, 246 had gender information and 199 had age information. Their mean age was 42.8 years, with standard deviation of 16.4 years. They were 64.2\% male. 

We also have information about disease severity in some COVID-19 patients. We have survival information about 81 patients, with a survival rate of 81.5\%. We have information about intubation on 69 patients, with an intubation rate of 59.4\%. We have information about supplemental oxygen on 94 patients, with 59.6\% of them needing it. At last, we have ICU information on 112 patients, 59.8\% of them were on the intensive care unit.

\section{Transfer Learning and Twice Transfer Learning}

\subsection{Introduction to transfer learning}
When we add dimensions to a neural network input, the data tends to become sparser. Thus, with larger inputs (like our 224x224 images), we need more data to create a good statistical model of the inputs and labels distribution. This problem is known as ``curse of dimensionality'' (\cite{Curse}).

Deep neural networks are mathematical models with many trainable parameters, enabling them to model complex data distributions and statistical relations. But, when we do not have enough data, this also makes them prone to learn small variations and noise in the training dataset, which are exclusive to that database and do not reflect the real phenomenon we are trying to model. Thus, with insufficient data, we can generate overfitting, hence creating a neural network that performs well on the training dataset but badly on the test database (\cite{goodfellow2016}). 

In summary, DNNs have a tendency to overfit when trained on small datasets and large inputs. Transfer learning is a technique that helps avoiding this problem. It consists in using a network that was already trained to solve a task in one dataset, and training it again (or fine-tuning) on another database, to solve another task. Doing this, we hope that representations learned by the DNN in the first database can help the model generalization on the second. This is particularly helpful when the first dataset is much larger than the second one (\cite{goodfellow2016}).

When we train a deep neural network, each layer learns to map the information it receives onto a new representation of the input data, creating what is called representation learning (\cite{representation}). Thus, the layers implicitly extract features of the inputs. The nearer from the network output a layer is, the higher the level of abstraction the learned feature has (\cite{goodfellow2016}). What makes transfer learning effective is that some features, learned from the first task and dataset, can help the DNN solve the second task, in the second dataset.  For example, a network trained with a large image classification dataset, like ImageNet (\cite{imagenet}), can learn, in that database, to identify image edges, a feature which can also be useful for interpreting X-rays, in a dataset like ChestXray-14.

We can observe that, if the two tasks are similar, more features learned in the first dataset will be useful in the second one, increasing the benefit of transfer learning. Thus, an ideal case would be to have a first dataset that is very large and whose task is very similar to that of the second one.

When we use an already trained network in another dataset, we need to pay attention to whether the input size remains the same (if not, the inputs are generally re-scaled). Also, because we are changing the task, the DNN output needs to change. One can add new layers at the end of the network or replace the last layers with new ones (\cite{goodfellow2016}). A common approach is to replace just the output layer,  removing it and adding a new one, with randomized weights and biases. The more similar the two tasks are, the more the last layers learned representations will be useful in the second dataset, and the more we would want to keep them. 

\subsection{Twice transfer learning}

It is common to choose ImageNet as the first dataset (\cite{chexnet}) when we have image classification tasks, as it is a database with millions of images and many classes. But one can argue that classifying these images is not a task particularly similar to that of classifying chest X-rays as COVID-19, pneumonia or normal.

NIH ChestX-ray14 classification was a task much more alike ours, and the dataset is still very large, with over 100.000 X-rays. Also, beginning with a DNN already pretrained on ImageNet would accelerate training on the NIH database and the network could keep some information, learned in ImageNet, through training on ChestX-ray14. This information might also help in the final fine-tuning, on the COVID-19 dataset. 

Thus, a twice transfer learning, or transfer learning in three steps seemed like a good proposition: a DNN would be first trained on ImageNet (\cite{imagenet}), then on ChestX-ray14 (\cite{NIHSet}) and, finally, on the COVID-19 dataset that we assembled. Fine-tuning CheXNet in the COVID-19 dataset indirectly created a transfer learning in three steps: we took a DNN that had already been trained on ImageNet and then on ChestX-ray14, and we applied the third step, training it on the COVID-19 dataset. 

But we can also train other neural networks with this twice transfer learning if, after downloading them pretrained on ImageNet, we train them on ChestX-ray14 and then on the COVID-19 dataset. Looking for twice transfer learning on other works, we found that it was already used, with success, for mammogram images (\cite{DoubleMamog}).

\section{Output neuron keeping}
In this paper, we propose an original addition to the twice transfer learning technique: output neuron keeping. In three-step transfer learning, we look for a task in the second step that is very similar to the final step task. Having two alike datasets, one might find that they share classes. In our study, ChestX-ray14 and the COVID-19 dataset have both a class for healthy individuals (called ``no findings'' in ChestX-ray14 and ``normal'' in the COVID-19 dataset). Also, ChestX-ray14 shares the pneumonia class with our dataset. 

Thus, we suggest that, having the same or very similar classes in the second and third step of twice transfer learning, when preparing the network for the third step, one could keep the output neurons that classify those classes and change only the other output neurons. Doing this, the representations that these artificial neurons learned in the second step can be maintained and this may improve training speed or performance in the final task. 

To keep output neurons, a simple approach in PyTorch begins by copying their weights and biases at the end of twice transfer learning step two (training on the second dataset). Then, in the beginning of step 3, change the DNN output layer to match the new desired output format, find the neurons that will classify the classes similar to step two's, and substitute their weights and biases for the copied ones.
 
\section{Trained DNNs}

In this work we trained five DNNs, which we will call A, B, C, D and E.

Network A is a 201 layers DenseNet, downloaded pretrained on ImageNet, and trained again on the COVID-19 dataset. Thus, it received a simple transfer learning approach.

Network B is also a DenseNet201, downloaded pretrained on ImageNet. But it was then trained on ChestX-ray14 and then on the COVID-19 dataset. Thus, it used twice transfer learning (with ImageNet in the first step, ChestX-ray14 in the middle step  and the COVID-19 dataset in the last).

Network C is the same as network B, but, besides the twice transfer learning approach, we used output neuron keeping: the neurons that classified the no findings and pneumonia classes in ChestX-ray14 were assigned to classify normal and pneumonia in the COVID-19 dataset, the other output neurons were removed and a new one, with random weights and biases, was added to classify the chance of COVID-19. Thus, here we used a twice transfer learning with output neuron keeping.

Network D is a 121-layer DenseNet. We downloaded a pretrained CheXNet (already trained on ImageNet and then on ChestX-ray14) and trained it on the COVID-19 database. So, it had a twice transfer learning, but only the last step was done by us.

The last network, E, is also a DenseNet121. It began as a pretrained CheXNet and we again trained it on the COVID-19 dataset. But, before training on this dataset, we copied the weights and bias from the neuron that classified pneumonia to the one that would classify pneumonia. Thus, it used a twice transfer learning with output neuron keeping, and maintained just one output neuron. We note that CheXNet had no neuron to classify the chances of healthy lungs to keep (it had 14 output neurons, one for each of the 14 diseases on ChestX-ray14).
 
Our motivation to choose working with dense neural networks was CheXNet excellent result in ChestX-ray14, even surpassing 4 radiologists in pneumonia detection (\cite{chexnet}). 
 
Table 1 summarizes all DNNs we created.

\begin{table}[]
\centering
\begin{tabular}{|l|l|l|}
\hline
Name & DNN architecture & Training process \\ \hline
A & 201-layers DenseNet & Transfer learning \\ \hline
B & 201-layers DenseNet & Twice transfer learning \\ \hline
C & 201-layers DenseNet & \begin{tabular}[c]{@{}l@{}}Twice transfer learning +\\ output neuron keeping\end{tabular} \\ \hline
D & 121-layers DenseNet (CheXNet) & Twice transfer learning \\ \hline
E & 121-layers DenseNet (CheXNet) & \begin{tabular}[c]{@{}l@{}}Twice transfer learning +\\ output neuron keeping\end{tabular} \\ \hline
\end{tabular}
\caption{DNNs}
\label{tab:my-table}
\end{table}

\section{Data Processing and Augmentation}

\subsection{ChestX-ray14}
As we would also train DenseNets in this dataset, we based our dataset processing for ChestX-ray14 in what the authors did when training CheXNet (\cite{chexnet}). All images were resized to 224x224 size, with 3 channels, and normalized with the mean and standard deviation from the network's previous training in ImageNet. We used the originally reported test dataset as our test dataset. The other images were randomly split, with 80\% of the X-rays to create the training dataset and 20\% for validation (holdout). Different datasets had no images from the same patient. 15 labels were created, one for each disease and one for ``no findings'' (they were organized in a binary vector with 15 dimensions).

Like the authors did when training CheXNet (\cite{chexnet}), we applied random horizontal flips (with 50\% chance) to the training images before giving them to the DNN. This was done online and, if the image was flipped, we would only feed the new image to the network, not the new and the old one (thus, not making the batch bigger).

\subsection{COVID-19}
Firstly, we divided our assembled COVID-19 dataset into three: training, validation and test. To create the test dataset, we randomly took 50 images of each class (normal, pneumonia and COVID-19).  After removing the 150 test images, we took 90\% of the remaining images for training and 10\% for validating.  This was also done randomly, but preserving the same class proportions in the two datasets. Again, different datasets had no images from the same patient. All images were loaded in greyscale (to minimize color variations between the datasets), reshaped from 1024x1024 pixels to 224x224, converted to three channels and then normalized to the same mean and standard deviation used in the ChestX-ray14 and ImageNet normalization.

 The CheXpert database had 8 pneumonia images that were labeled by three board-certified radiologists. These images were included in our testing dataset, along with 42 other random pneumonia images from the CheXpert.

Many of the images had letters or words on them, and some of these words were exclusive for certain classes. For example, some COVID-19 images (from Italy) had the word ``SEDUTO'' (Italian word for ``seated") written on the upper left corner. We were afraid that this could affect the network classification performance, hence we decided to manually edit our test dataset images, removing the words or letters. They were simply covered with black rectangles and, as they were not over the lungs, no relevant information was lost. The objective was only to test the network ability analyzing the lungs and, by editing only the test dataset, there would be no risk of teaching the DNN to identify our black rectangles during training.

We decided to apply data augmentation for two reasons: it improves the DNN performance for small datasets (like our COVID-19 database), and because it would balance our datasets. As we would also benefit from a balanced validation dataset, we applied augmentation in training and validation. 

We used three image augmentation methods: rotations (between -40 and 40 degrees), translations (up to 28 pixels left and right or up and down) and flipping (horizontal). These transformations could augment our data and also make the DNN more robust to input translations and rotations. All augmentation was done online and, after one operation, we would not substitute the original image, we would just add the new one, randomly rotated, translated and possibly flipped (50\% chance), to the batch (making it bigger). We augmented our normal image database 30 times, our pneumonia images 8 times, and our COVID-19 images 24 times. We ended up with a training dataset of 8280 COVID-19, 8640 pneumonia and 8640 normal lung augmented images. To feed the DNN a completely balanced dataset, 360 pneumonia and normal lung augmented images were left out in each epoch. In every epoch we randomly changed which images would be left out, therefore every image was used in the training process.

\section{Creating and Training the DNNs}

\subsection{On the ChestX-ray14 dataset}
As networks B and C have the same training process in the twice transfer learning first and second steps, we are able to train only one network on the ChestX-ray14 database, which would be used for creating networks B and C in the future.

To create the DNN we downloaded a pretrained PyTorch version of DenseNet201. The only changes we made on it was substituting its output layer for one with 15 neurons (one for each of the 14 diseases in this dataset and one for the ``no findings'' class) and we kept this layer's activation as a sigmoid function. The training process was carried out in PyTorch, with binary cross entropy loss, stochastic gradient descent with momentum of 0.9, mini-batches of 16 images and hold-out validation. We trained the networks on two NVidia GTX 1080 GPUs.

We began by freezing all model parameters except for the output layer's and training for 20 epochs, with a learning rate of 0.001. We then set the learning rate to 0.0001, unfroze all the model parameters and trained for 90 epochs more, in the end of which the DNN was already overfitting.

\subsection{On the COVID-19 dataset}
To create network A we downloaded the ImageNet pretrained PyTorch version of DenseNet201, removed the output layer and added a new one, with three neurons and softmax activation. For DNN B we started with the neural network we had trained on ChestX-ray14, also removed its last layer and added a new one, like the above mentioned. For network C we made the same output layer substitution in the network we had trained on ChestX-ray14, but we copied the weights and biases from the output neurons that classified ``no findings'' and ``pneumonia'' to the ones that would classify ``normal'' and ``pneumonia'' in the COVID-19 dataset. 

For network D, we downloaded a pretrained CheXNet (on ImageNet and ChestX-ray14) from \cite{pretrained}, and proceeded by also changing its final layer for one with three neurons and softmax activation. Finally, for network E, we downloaded the same network (\cite{pretrained}), also changed the final layer as in network D, but we copied the weights and bias for the output neuron that classified pneumonia to the one that would again classify pneumonia, in the new dataset.

For all networks, the training process in the COVID-19 dataset was the same, given that their architectures were similar, and this allowed us to better compare the transfer learning methods. We used PyTorch, cross entropy loss, stochastic gradient descent with momentum of 0.9 and mini-batches of 9 images. We trained the networks on two NVidia GTX 1080 GPUs. We also used holdout validation. Most training parameters were determined with many preliminary tests, for example, weight decay was used in the beginning to avoid fast overfitting, but was removed when we noticed the DNNs had stopped improving training error.

The training process had 4 phases, which we will describe now. We began by freezing all the network parameters except for the output layer's and we trained for 10 epochs, using a learning rate of 0.001 and weight decay of 0.01. We then unfroze all parameters, trained for 48 epochs, with early stop and patience of 20, learning rate of 0.0001 in the last layer and  decreasing by a factor of 10 for each previous dense block and its transition layer. We lowered the learning rate to 0.00001 (now in every layer) and trained for 48 epochs more, with the same early stop and weight decay. Finally, for the last phase, we removed the weight decay and early stop and trained for 48 epochs again.

\section{Layer-wise Relevance Propagation}

Layer-wise Relevance Propagation (LRP) is an explanation technique that aims to make DNNs (complex and nonlinear structures with millions of parameters and connections) interpretable by humans. It decomposes the network prediction, showing, in a heatmap, how each input variable contributed to the output (\cite{LRP}). We note that interpreting deep neural networks can be challenging. A Taylor Decomposition, based on the Taylor expansion of the network output, is unstable in deep neural networks, due to noisy gradients and the existence of adversarial examples (\cite{LRPBook}).

We may choose any output neuron to start the relevance propagation, and this choice will define the meaning of the colors in the resulting heatmap. For example, if we choose to start LRP on the output neuron that classifies COVID-19, red colors on the map will indicate areas that the DNN associated with COVID-19, while blue areas will point characteristics associated with the other classes, normal and pneumonia. Examples can be seen in the section ``Analysis with LRP''. Unless stated otherwise, the heatmaps shown in this paper (section ``Analysis with LRP'') were created starting the propagation at the winning output neuron, i.e., the one with the highest output for the input X-ray.

LRP is based on propagating the DNN prediction backward through the layers, using local propagation rules, which may change for different layers. This relevance propagation has a conservation property, in the way that the quantity of relevance a neuron receives from the upper layer will be distributed in equal amount to the neurons in the lower layer (\cite{LRPBook}). This property ensures that the quantity of explanation we get in the input (in the heatmap) relates to what can be explained by the output. As examples of medical contexts in which LRP was used we can cite neuroimaging (\cite{fmri-lrp}) and explaining therapy predictions (\cite{LRPMedical}). 

Analyzing our network with LRP allows us to identify problems in the DNN classification method, and also to generate a heatmap of the X-ray image, showing where in the lungs the network identified issues. This map could be given to radiologists along with the network predictions, helping them to verify the classifier analysis, providing insights about the X-rays and allowing a more profitable cooperation between human experts and artificial intelligence.

 We can choose between many propagation rules in each neural network layer, and presets are selections of these rules for the many layers in a DNN. We can compare them, searching for one that creates good human interpretability and fidelity to the network operation. We used the Python library iNNvestigate (\cite{innvestigate}), which already implemented LRP for DNNs like DenseNet and has parameter presets that work well for these networks. This library works with Keras and TensorFlow, but we trained our DNNs on PyTorch, thus, we used the library pytorch2keras (\cite{py2keras}) to convert our models. After the conversion, we tested them again, and obtained the same accuracies we had on PyTorch, confirming that the conversion worked well.

\section{Results}
\label{}

Figure 1 shows how test accuracy changed during training on the COVID-19 dataset, for all five DNNs. These measurements were taken for the best networks (according to validation loss) in each of the four training phases described in section ``Creating and Training the DNNs'' (which correspond to epochs 10, 58, 106 and 154). Our best test accuracy was 100\%, achieved by both the CheXNets and the DenseNet201 with twice transfer learning and output neuron keeping. 
\begin{figure}[t]
\caption{Test accuracies during training plot.}
\includegraphics[width=1\textwidth]{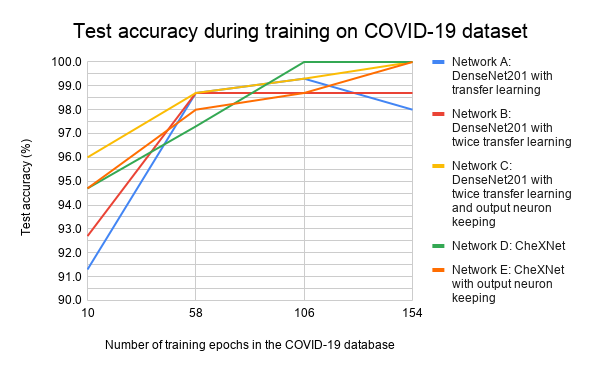}
\centering
\end{figure}

 In Table 2 and 3 we show the confusion matrices from networks that did not achieve 100\% accuracies, DNNs A and B. Network B made 2 mistakes in the 150 test images: 2 COVID-19 images were classified as pneumonia. DNN A misclassified 1 healthy patient as COVID-19.

With 100\% accuracy, our DNNs C, D and E have precision, recall and F1 score of 1 in our test dataset.

\begin{table}[]
\centering
\begin{tabular}{ll|l|l|l|}
\cline{3-5}
 &  & \multicolumn{3}{l|}{Predicted class} \\ \cline{3-5} 
 &  & Normal & \begin{tabular}[c]{@{}l@{}} Pneumonia\end{tabular} & COVID-19 \\ \hline
\multicolumn{1}{|l|}{\multirow{3}{*}{Real Class}} & Normal & 49 & 0 & 1 \\ \cline{2-5} 
\multicolumn{1}{|l|}{} & \begin{tabular}[c]{@{}l@{}} Pneumonia\end{tabular} & 0 & 50 & 0 \\ \cline{2-5} 
\multicolumn{1}{|l|}{} & COVID-19 & 0 & 0 &50 \\ \hline
\end{tabular}
\caption{Confusion matrix for network A}
\label{tab:my-table}
\end{table}

\begin{table}[]
\centering
\begin{tabular}{ll|l|l|l|}
\cline{3-5}
 &  & \multicolumn{3}{l|}{Predicted class} \\ \cline{3-5} 
 &  & Normal & \begin{tabular}[c]{@{}l@{}} Pneumonia\end{tabular} & COVID-19 \\ \hline
\multicolumn{1}{|l|}{\multirow{3}{*}{Real Class}} & Normal & 50 & 0 & 0 \\ \cline{2-5} 
\multicolumn{1}{|l|}{} & \begin{tabular}[c]{@{}l@{}} Pneumonia\end{tabular} & 0 & 50 & 0 \\ \cline{2-5} 
\multicolumn{1}{|l|}{} & COVID-19 & 0 & 2 & 48 \\ \hline
\end{tabular}
\caption{Confusion matrix for network B}
\label{tab:my-table}
\end{table}

We applied Layer-wise Relevance Propagation to the trained neural networks. The generated heatmaps allowed us to analyze how each part of the input X-rays influenced the DNN classification. This topic will be discussed in more detail in section ``Analysis with LRP'', along with examples of the heatmaps.

\section{Discussion}

Analyzing Figure 1, we start comparing the DenseNet201 DNNs. In the first epochs, we note that network C, with twice transfer learning and output neuron keeping, started with noticeably better accuracy, followed by network B and then network C, which used simple transfer learning. The performance differences become smaller with more training. When we compare the best results, we observe that DNN C had 100\% accuracy, DNN A 99.3\% and DNN B 98.7\%. 

Comparing the two CheXNets, we observe that their performances were similar, both had the same accuracies in the beginning and achieved equal results in the end. We think that output neuron keeping with only one neuron had a smaller effect than with two neurons, therefore the difference between networks B and C are larger than between DNNs D and E. We also see that these networks started with accuracies higher then network A and B, but smaller then DNN C, which used output neuron keeping on two neurons. 

 Analyzing tables 2 and 3 we observe that even on the networks that committed mistakes, no patient with some disease was classified as healthy, which would be the most dangerous type of mislassification.

We achieved test accuracies of 100\% with three networks, but we must note that our test dataset only has 150 images. Maybe these DNNs could make some mistakes in a bigger test database (which we did not use due to the limitation in the number of COVID-19 images).

Looking at the state-of-the-art in COVID-19 detection with deep learning, we observe that, according to the review \cite{review}, most of the techniques using DNNs have accuracies in the 90\% to 100\% range. Therefore, our work is on par with the current state-of-the-art.

\subsection{Analysis with LRP}

We tested different LRP presets on iNNvestigate and got more understandable and coherent heatmaps with ``LRP-PresetAFlat''. Figure 2 shows a correctly classified COVID-19 X-ray test image and the heatmap for it, taken from one of our best performing DNNs, network C. The more red the region on the map, the more important it was for the DNN classification as COVID-19. The more blue, the more that region is related to other classes (like a healthy part of the lung). We observed our DNN found signs of COVID-19 in both lungs. We also note that, on some images, our black rectangles can create some artifacts in their borders (red or blue lines, as can be seen on the neck region of figure 4), but, given that they were used only for testing and in all classes, we do not think this can affect accuracy.

\begin{figure}[t]
\caption{Heatmap for test COVID-19 image}
\includegraphics[width=0.8\textwidth]{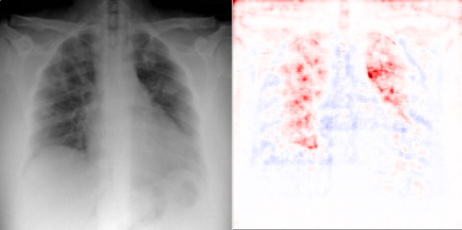}
\centering
\end{figure}

We decided to analyze the effect of words and letters on the X-ray images, fearing that words used on the datasets could be sources of bias. We used the same test COVID-19 image shown in figure 2, but without removing the word ``SEDUTO'' from its upper right corner and the letters ``DX'' from the upper left corner. The resulting heatmap is shown in Figure 3 (also created with network C). It becomes clear, by the red color on the map, that the network learned to associate these words with the COVID-19 class. 

\begin{figure}[t]
\caption{Heatmap for test COVID-19 image without removing word ``SEDUTO" and letters ``DX''}
\includegraphics[width=0.8\textwidth]{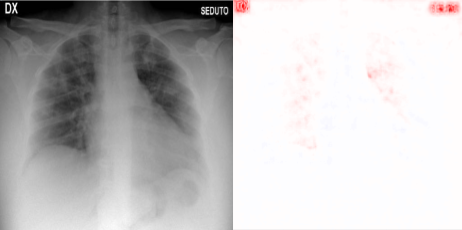}
\centering
\end{figure}

To measure the effect of this problem we tested the DNNs we trained with our testing dataset but unedited (with the words and letters that it originally had). This test generated only small changes in accuracy, which increased or decreased, at most, 1.33\% on fully trained networks. DNNs on early training stages showed changes up to 3.33\%. Our best neural networks (C, D and E), when fully trained, showed no accuracy change.

Another test was trying to ``fool'' our networks, adding to a COVID-19 lungs X-ray test image the letters ``L PA'', which were copied from a healthy image (L above and PA below, in a small rectangle). The network C given probability for COVID-19 changed from 99.94\% to 99.91\%, and for normal increased from 0.036\% to 0.055\%. Figure 4 shows a heatmap created by LRP starting on the output neuron that classifies the normal class. Therefore, red colors indicate areas associated with normal. We can observe that the letters PA are red on the heatmap, meaning that they are influencing the DNN to choose normal instead of COVID-19. The letter L, above the PA, is blue. This letter was very common in all classes, thus it was not associated with the normal class. Given the small output probabilities change, we see that, in this case, the letters effect was tiny.

\begin{figure}[t]
\caption{Normal output neuron heatmap for test COVID-19 image edited with letters (indicated by arrows) copied from a normal image}
\includegraphics[width=0.8\textwidth]{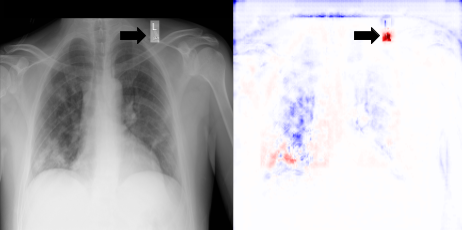}
\centering
\end{figure}

\section{Conclusion}
\label{}

The proposed method of output neuron keeping, with twice transfer learning,  outperformed the sole use of twice transfer learning and simple transfer learning in the 201-layer dense networks. Taking into account this work and the great results three steps transfer learning had in \cite{DoubleMamog}, we think that the technique and our output neuron keeping method are promising and could also improve performances in other classification problems. 

We were surprised by the fact that the CheXNet DNNs could keep up with and even surpass most of the DenseNet201 DNNs. We concluded that the main reason for this is that, even though we used data augmentation, the effect of overfitting was stronger on the 201-layer dense neural networks. 

LRP showed promising results highlighting details in the X-rays that most influenced the network classification. We hope that this may indicate a possibility to help radiologists and provide a better interaction between experts and artificial intelligence. It also allowed us to discover that words and letters can influence the DNN classifications. This influence was small in the fully trained DNNs, with accuracies changing at most 1.33\% if we do not remove the words and letters from the testing dataset. On DNNs at the beginning of the training process this effect was larger, with accuracies increasing or decreasing up to 3.33\%.

We should state that the COVID-19 dataset used in this study is not ideal. Although we used the largest open COVID-19 X-ray database that we could find in October 2020, we could only utilize 439 coronavirus X-rays. A much larger dataset, with all the classes collected from the same sources, would allow the creation of better generalizing classification models. It would also reduce possible causes of bias, as different sources can have different characteristics, like the letters and words that we analyzed in this work. We emphasize the need for such a database and hope to continue this research when it becomes available.  Furthermore, clinical studies would be required to ensure that the high accuracies obtained in this and other studies (\cite{review}) would also be achieved in a real world scenario.

Although larger databases and clinical studies are still needed, this study and other initiatives (\cite{CovidNet}, \cite{review}) show that DNNs have the potential of making chest X-ray a fast, accurate, cheap and easily available auxiliary method for COVID-19 diagnosis. The trained networks proposed here are open source and available for download in \cite{MyNets}: we hope DNNs can be further tested in clinical studies and help in the creation of tools to fight the COVID-19 pandemic.

\section{Acknowledgments}
This work was partially supported by CNPq (process 308811/2019-4) and CAPES.

\bibliographystyle{elsarticle-harv} 
\bibliography{mybibfile}

\end{document}